\documentclass{elsart}%
\usepackage{amssymb}
\usepackage{amsmath}
\usepackage{amsfonts}
\usepackage{graphicx}%
\begin{document}
\begin{frontmatter}
\title{On X-Ray Waveguiding in Nanochannels: Channeling Formalism}
\author{S.B.~Dabagov \corauthref{*}}
\ead{dabagov@lnf.infn.it} \corauth [*]{Corresponding author. Tel.:
+39 06 9403 2877; Fax: +39 06 9403 2427}
\address{INFN - Laboratori Nazionali di Frascati,  Via E. Fermi 40, \\
I-00044 Frascati, Italy}
\address{RAS - P.N. Lebedev Physical Institute, Leninsky pr. 53, \\
119991 Moscow, Russia}
\begin{abstract}
The question on X-ray extreme focusing (smallest reachable spot
size) brings us to the idea for using the wave features of X-ray
propagation in media. As known, wave features are revealed at
propagation in ultra-narrow collimators as well as at glancing
reflection from smooth flat and/or strongly curved surfaces. All these
phenomena can be described within the general formalism of X-ray
channeling.
\end{abstract}
\end{frontmatter}


\section{Introduction}

\label{intro}

The advent of the nanotechnology era has given rise to unprecedented
opportunities, challenges and enormous potential for new products and wealth
creation. Manipulation of matter on the atomic scale will require new tools
for lithography and metrology and radiation at X-ray wavelengths will be
fundamental to the development of this area. Progress within the fields of
nanotechnology and X-ray propagation will be mutually beneficial. For
instance, some types of processes based on the self-organization of materials
that have recently attracted considerable interest because of the possibility
of preparing fine patterns of nanometer dimensions over larger areas, can be
used for the fabrication of X-ray waveguides
\cite{spiseg-apl1974,pogossian-opcom1995}. Among them the formation of
highly-ordered aligned carbon nanotubes and ordered arrays of uniform-sized
porous in anodic alumina is of great interest. X-ray propagation in
$n$-channels\footnote{ Below in the text we use $n$- for nano- and $\mu$- for
micro-.} is important due to potential applications in X-ray optics. A special
feature of these structures is a long, hollow, inner cavity, which could act
as a channel for selective radiation penetration, similar to channeling of
charged particles in crystals (see \cite{da-ufn2003} and Refs. in);
$n$-channels can be considered as capillaries (the base of
capillary/polycapillary optical elements \cite{kuko-phrep1990,Engstrom,Stern}%
). Research in X-ray propagation in capillary structures shows that
diminishing the capillary internal radius from microns to nanometers results
in a change of the character of radiation propagation, from the surface
channeling in $\mu$-capillaries down to bulk channeling in $n$-capillaries .
Numerical simulations \cite{zhevago,dedkov} have shown that carbon nanotubes
will act as soft X-ray waveguides and support modes of propagation when coated
with various materials \cite{childs...phe2003}. Other researchers have already
demonstrated experimentally the coherent propagation of X-rays in a planar
waveguide with a tunable air gap \cite{zwanenburg...prl1999}. The angular
dependence of the intensity of C K$_{\alpha}$ radiation vs the aligned carbon
nanotubes' orientation suggests the possibility of X-ray channeling as well as
radiation diffraction on nanotube multiwalls \cite{okotrub...jetpl2005}.

It is important to clarify the origin of some peculiarites just from the
beginning in order to prevent misunderstanding (that time by time takes place
in publications) in the processes observed or to be expected. As shown below,
one exists a well specified difference in the origin of trapped (bound states)
radiation propagation for $\mu$-- and $n$--size channel
structures\footnote{The latter is mainly related to propagation of X-rays in
capillary structures (mono- and polycapillaries of various origins and sizes,
and optical elements on its base).}.

The main criterion for observing the wave features of radiation propagation in
media of the length $L$ is very simple - the transverse space where radiation
is limited at propagation, does not matter by either a profiled surface or a
specific collimation system, should approach by size the transverse radiation
wavelength: $X_{\perp}\simeq\lambda_{\perp}$ ($\lambda_{\perp}\simeq
\lambda/\vartheta_{c}$, where $\vartheta_{c}=\omega_{p}/\omega$ with $\omega$
as photon energy and $\omega_{p}=\sqrt{4\pi n_{e}e^{2}/m_{0}}$ - the plasmon
energy, $n_{e}$ is the electron density of the cladding, $e$ and $m_{0}$ are
the electron charge and mass). For X-ray frequencies, at reflection from a
flat surface (optimally, we deal with the total external reflection) as well
as at X-ray propagation in a planar waveguide, the transverse dimension of a
beam can be estimated as $X_{\perp}\simeq L\vartheta_{c}$; and thus we get
very simple and important expression that allows the limits for revealing the
wave features at reflection to be evaluated%
\begin{equation}
L\vartheta_{c}^{2}\simeq\lambda\label{wave criterium}%
\end{equation}
It makes evident why even at radiation propagation in $\mu$-channels\ it is
possible to observe the wave features. This relation will be in details
examined below from the view point of the both wave equation solution and
simple physical base of the phenomenon.

The modes of radiation propagation in a waveguide are revealed at interference
between the incident and reflected waves forming a standing wave pattern
\cite{spiseg-apl1974}. However, it becomes constructive just for specific
angles. This phenomenon, valid for reflection from a flat surface, takes place
just in the vicinity of the surface. Similar phenomenon can be observed at
radiation \ reflection from the curved surface (so called "whispering modes")
\cite{vino1}. Strong radiation redistribution takes also place behind
capillary systems (which is actually a simple example of the curved surface
system); some structural features in the distribution are due to the spatial
geometry of the system (typically, hexagon type in the transverse cross
section). However, some fine features could not be interpreted by the ray
optics, and require solution of the wave equation of radiation propagation.

History of interference phenomena observed for X-rays, from various sources,
propagating in capillary structures counts more than 10 years. First
theoretical note regarding such a possibility for capillary optical systems,
where the X-ray interference behind capillaries, was published as an internal
note \cite{da-rep92}. After that, during 1993-1994 this phenomenon was
observed in a set of experiments with polycapillary optical elements (channels
sizes of $\ \sim100$ $%
\mu
$m) in the beams of synchrotron radiation ($\lambda\simeq$ 10 \AA ,\ S-60
LPI). Then, the first joint paper appeared \cite{da...jsr95} with the wave
interpretation of the features recorded. Phenomenology of the phenomenon given
in \cite{da...pla1995} has shown that the fine features of X-ray propagation
in $\mu$-size channels, which have been observed behind capillary structures,
can be explained in view of the radiation interference due to the various
channel's curvatures. Later the trapped radiation propagation in the very
vicinity of a surface was carefully studied in a number of works
\cite{daku-spie95,alda-nim98,arar-phscr98,coh-incoh,liu...prl97,kukh.etal-nimb00,ca-da...apl01,dabagov-xrs2003}
where\ the wave theory of X radiation propagation along a curved surface was
developed (for complete citation, please, see Refs. in \cite{da-ufn2003}).

Regardless of the research on radiation transmission by capillary systems,
which represent circular guide systems, nowadays, considerable progress in
studying of X-ray waveguiding in the planar structures (as specially
fabricated waveguides consisting of the guiding and cladding layers, or
\ ultra narrow slits and collimators with air gap, etc.) is achieved
\cite{zwanenburg...prl1999,zwanenburg...prl2000,pfeiffer...science2002,bongaerts...jsr2002,bergemann...prl2003,jarre...prl2005,fuhse.salditt-phis.b2005,fuhse.salditt-optcomm2006,bukreeva...prl2006}%
. Additional to the planar case, in paper \cite{bergemann...prl2003} the
authors have proposed a wave theory of X-ray propagation through a circular
guide. It is notable for this paper that the wave theory of X radiation
propagation in capillaries (circular shaped guides!) was developed in many
previous papers as a general theory of radiation propagation for both $\mu
$-size and $n$-size capillaries. Detailed analysis presented in many articles
is based on the transverse wavelength aproach as a main criterion (see the
review  \cite{da-ufn2003}). The authors of the paper
\cite{bergemann...prl2003}, thanking D. Bilderback for the information on
possibility to observe an interference picture behind capillary/polycapillary
systems, have forgotten to make any reference to previously published results;
the adequate citation were done just for the case of planar
waveguides\footnote{Moreover, in the paper there is no citation to the
Bilderback's paper \cite{bilderback-xrs2003} published as a review on the
discussions at the ICOM'2001 meeting (International Capillary Optics Meeting,
Antwerp, 17-21 June 2001), where the wave features of X radiation propagation
in capillary systems, theoretically evaluated and experimentally proved, were
presented and discussed. These results were published in the same volume of
X-Ray Spectrometry issued as a special volume of the ICOM'2001 proceedings
\cite{dabagov-xrs2003,kukhlevsky-xrs2003}.}. Since 1994 new coherent features
of X radiation propagation in capillary structures have been discussed at many
meetings: SPIE annual conferences, specially organized meetings on capillary
optics and its applications (International Capillary Optics Meeting - Antwerp
2001; International Conference on Capillary X-Ray and Neutron Optics -
Zvenigorod 2001 and 2003, International Conference on Charged and Neutral
Particles Channeling Phenomena \ - "Channeling 2004 " and \ "Channeling 2006 "
Frascati), practically all conferences on X-ray physics.

In this work after the introduction to physics of coherent phenomena of
radiation scattering in $\mu$-- and $n$-structures and X-ray channeling in
hollow channels of various origins, the results obtained within international
collaborations will be presented.

\section{General theory of X-ray channeling}

As shown recently, the propagation of X-ray photons through the narrow guides
exhibits a rather complex character
\cite{da...jsr95,bongaerts...jsr2002,bergemann...prl2003}. Not all the
features shown experimentally can be explained within the geometrical (ray)
optics approximation \cite{arar-phscr98,kukh.etal-nimb00,mono}. On the
contrary, application of the wave optics methods allows the processes of
radiation transmission by the guides to be described in details.

\noindent The passage of X radiation through the guides is mainly defined by
its interaction with the inner guide walls. In the ideal case, when the
boundary between hollow channels and walls represents a smooth edge, the beam
is split in two components: the mirror-reflected and refracted ones. The
latter appears sharply suppressed in the case of total external reflection.
The characteristics of scattering inside the structures of ultra-small holes
of various shapes can be evaluated from solution of the Helmholtz equation. In
the first order approximation, propagation of X radiation through specially
designed guides, the cladding material of which is characterized by the
refractive index $n=1-\delta(\mathbf{r})+i\beta(\mathbf{r})$ defining by the
guide geometry, is described by a wave propagation equation%
\begin{equation}
\left(  \mathbf{\Delta}+k^{2}n^{2}(\mathbf{r})\right)  E(\mathbf{r}%
)=0,\ n\equiv\left\{
\begin{array}
[c]{rcl}%
1 & , & \text{hollow core}\\
n_{0}=1-\delta_{0}+i\beta_{0} & , & \text{cladding}%
\end{array}
\right.  \label{helmholtz}%
\end{equation}
for the electromagnetic field amplitude $E$, where $\mathbf{k\equiv
(}k_{\parallel},k_{\perp}\mathbf{)}$ is the wave vector of radiation,
$k=2\pi/\lambda$ , $\mathbf{\Delta}\equiv\partial^{2}/\partial\mathbf{r}%
_{\perp}^{2}+\partial^{2}/\partial z^{2}$ is the Laplasian, $\mathbf{r}%
\equiv(\mathbf{r}_{\perp},z)$. Separating a transverse part of the radiation
field as $E(\mathbf{r})=E(\mathbf{r}_{\perp})e^{ik_{\parallel}z}$, the
Helmholtz equation can be reduced to%
\begin{equation}
\mathbf{\nabla}_{\perp}^{2}E\left(  \mathbf{r}_{\perp}\right)  =\left(
2k^{2}\delta-k_{\perp}^{2}\right)  E\left(  \mathbf{r}_{\perp}\right)
\ ,\label{general wave}%
\end{equation}
\noindent where the right side term in brackets is a potential of interaction
$V_{eff}$. \noindent Due to the fact that the transverse wave vector
$k_{\perp}\approx k\vartheta$ under the grazing wave incidence ($\vartheta
\ll1$), an \textquotedblright effective interaction
potential\textquotedblright\ is estimated by the expression%
\begin{equation}
V_{eff}\left(  \mathbf{r}_{\perp}\right)  =k^{2}\left(  2\delta\left(
\mathbf{r}_{\perp}\right)  -\vartheta^{2}\right)  =\left\{
\begin{array}
[c]{rcl}%
-k^{2}\vartheta^{2} & , & \text{guiding channel}\\
k^{2}\left(  2\delta_{0}-\vartheta^{2}\right)   & , & \text{cladding}%
\end{array}
\right.  \label{e-pot-flat}%
\end{equation}
From the latter the phenomenon of total external reflection at $V_{eff}=0$
follows, when $\vartheta\equiv\vartheta_{c}\simeq\sqrt{2\delta_{0}}$ is the
Fresnel's angle. One can see that Eq.(\ref{general wave}) with the effective
potential (\ref{e-pot-flat}) corresponds to the Schr\"{o}dinger equation for a
massive particle motion, $\mathbf{\nabla}_{\perp}^{2}\equiv p^{2}/(2m)$, in a
specified potential well $V_{eff}$. That's why it becomes much more convenient
to use below the terminology of "channeling" \cite{Lindhard}, where a channel
is formed by the effective potential of radiation interaction in a guide (a
quantum well)\footnote{It is well known that channeling of charged particles
may take place when the small divergent beams are transversing crytals near
the main crystallographic planes or axes at the angles less than some critical
one $\varphi_{L}$ known as the Lindhard angle of channeling ($\varphi
_{L}\simeq\sqrt{V_{eff}/\varepsilon}$, $\varepsilon$ is the particle
energy).}. As well known from quantum mechanics, any well is able to support
at least one quantum bound state (channeling state); the number of the states
can be estimated from the expression for the potential (\ref{e-pot-flat}). The
equation (\ref{general wave}) for radiation propagation in a media with the
potential (\ref{e-pot-flat}) can be solved for the case of $\mu$-channels as
well as for the $n$-guides. It is important to underline the main difference
of the radiation propagation in $\mu$- and $n$-channels that is defined by the
ratio between the effective channel size and the transverse wavelength of
radiation. Moreover, in $\mu$-channels the main parameter of the guide is its
shape (collimation profile or surface curvature) whereas for $n$-channels it
is the transverse channel size.

However, there is another very interesting case to be examined, namely, when
the radiation is propagating along the curved surface (so called multiple
reflection regime). It is of strong importance in case of the circular $\mu$-guides.

\noindent When the reflecting surface is not more flat but curved, the
effective potential reveals an additional contribution. To describe new
expected features due to the new surface profile, we have to take into account
the fact that reflection of electromagnetic wave occurs on extended area. The
minimal size of it at sliding angles is defined by $\left(  \Delta d\right)
_{\parallel\min}\sim(4\pi c/\omega_{p})$ $\vartheta_{c}^{-1}$($c$ is the light
velocity) that is much greater than the atomic distances. Due to this fact, it
is possible to consider that interaction takes place in a macro field
characterized by a macroscopic dielectric permittivity $\varepsilon_{0}\equiv
n_{0}^{2}$. The last determines reflection and absorption characteristics of
interaction. Thus, the reflection coefficient is defined by polarization and
absorption in the layer of electric field penetration $(\Delta d)_{\perp}%
\sim2\pi c/\omega_{p}$. General analysis shows, that at sliding angles less
than the Fresnel's angle, TER is observed. Within the geometrical optics
appoximation, the coherent scattering occurs in a specular (mirror) direction,
while the incoherence is taken into account by the radiation absorption in the
reflecting layer. In other words, a new term in the potential of interaction
can be considsered as corresponding to the additional \textquotedblright
potential energy\textquotedblright. Due to the reflecting surface curvature a
photon \ "receives" an angular momentum $kr_{curv}\varphi$ , where $r_{curv}$
is a curvature radius of the photon trajectory. The latter is supplied by the
\textquotedblright centrifugal potential energy\textquotedblright%
\ $-2k^{2}r_{\perp}/\left(  r_{curv}\right)  $ \cite{liu...prl97}%
\begin{equation}
V_{eff}\left(  \mathbf{r}_{\perp}\right)  =k^{2}\left(  2\delta\left(
\mathbf{r}_{\perp}\right)  -\vartheta^{2}-2\frac{r_{\perp}}{r_{curv}}\right)
.\label{e-pot-curv}%
\end{equation}
\noindent Because of variation in the system spatial parameters, the
interaction potential has been changed from the step potential with the
potential barrier of $2k^{2}\delta_{0}$ to the well potential, with the depth
and width defined by the channel characteristics.

Above we have described a general rule to be fullfiled in order to reveal the
radiation wave behaviours. Obviously, it should be valid for any optical
scheme regardless both its size and shape (we can consider just two important
parameters of any device: guide channel and its shape, which is actually
defined by the interacting surface curvature). \noindent Let us now to
estimate the limit in the curvature radius $r_{curv}$, at which the wave
behaviours are displayed under propagation of radiation in channels
\cite{da-rep92} (in the case of capillaries or system of capillaries it is
defined by its channel diameter and bending). Let consider a photon with the
wave vector $\mathbf{k}$ propagating along a curved surface with the curvature
radius $r_{curv}$ (at the angles less than the critical one $\vartheta_{c}$,
it happens as surface channeling, the bound states will be evaluated below in
the text). At small glancing angles, $\vartheta<\vartheta_{c}$, the change of
the longitudinal wave vector, $k_{||}$, under reflection from a surface is
negligibly small; but one mainly changes the transverse wave vector
$k\mathbf{_{\perp}}$, $k_{\perp}\simeq k\vartheta_{c}\;(\vartheta_{c}\ll1)$.
\noindent Correspondingly, it follows that the transverse wavelength will much
exceed the longitudinal wavelength that provides the interference effects to
be observable even for very short wavelengths. Indeed, $\lambda_{\perp}%
\simeq\lambda/\vartheta_{c}>>\lambda$, and quantum mechanical principles
postulate that, in order to display the wave properties of a channeling
photon, it is necessary that typical trasnverse sizes of an \textquotedblright
effective space\textquotedblright\ $\delta_{i}$ , in which waves have been
propagating, be commensurable with the transverse wavelength, i.e. $\delta
_{i}(\vartheta)\equiv\lambda_{\perp}$. This condition may be rewritten in the
following form:%
\begin{equation}
r_{curv}\vartheta_{c}^{3}\simeq\lambda\label{b-cond}%
\end{equation}
\noindent So, from this simple estimate we can conclude that the relation
(\ref{b-cond}) provides a specific dependence for surface bound state
propagation of X-rays - \textit{surface channeling} - along the curved
surfaces (for instance, in capillary systems). Taking into account that in
approximation of the glancing angles $\vartheta\ll1$ one defines the
longitudinal size of the curved reflecting surface as $L\simeq r_{curv}%
\vartheta$, the expression (\ref{b-cond}) is in rather good agreement with a
general condition (\ref{wave criterium}) to be satisfied at any optical device
for observing wave character of radiation propagation.

In the following we briefly discuss a solution of the wave equation in the
case of an ideal reflecting surface (i.e. without roughness), when the
reflected beam is basically determined by the coherently scattered part of
radiation (for details see  \cite{coh-incoh}). Evaluating the wave equation
with the boundary conditions of a channel shows that X-radiation may be
distributed over the bound state modes defined by the channel potential. It is
important to note here that the channel potential acts as an effective
reflecting barrier, and then, the effective transmission of X-radiation by the
hollow guides is observed. While the main portion of radiation undergoes
incoherent diffuse scattering, the remaining contribution (usually small) is
due to coherent scattering that represents a special phenomenon, extremely
interesting to observe and clarify
\cite{da-ufn2003,bongaerts...jsr2002,kukhlevsky-ch2006report}.

\section{Surface channeling: $\mu$-channels}

\subsection{\noindent Planar guide}

As mentioned above, in $\mu$-channels the propagation features are defined by
the radiation interaction with a surface. Taking into account that a skin
layer, where the reflection is formed, is negligible thin in respect with the
guide wall thickness, the amplitude of the guiding wave can be presented as%
\begin{equation}
E_{m}(\mathbf{r})=u_{m}(\mathbf{r}_{\perp},z)\ e^{-ik_{\parallel m}z}\ ,
\label{general solution}%
\end{equation}
\noindent where $\mathbf{r}_{\perp}\equiv x\mathbf{e}_{x}$ for a planar, 1D
waveguide and $\mathbf{r}_{\perp}\equiv x\mathbf{e}_{x}+y\mathbf{e}_{y}$ for a
2D waveguide,\ $\left\vert \mathbf{e}_{x}\right\vert ^{2}=\left\vert
\mathbf{e}_{y}\right\vert ^{2}=1$, $k_{\parallel m}$ is the propagation
constant. Neglecting absorption, in case of a planar waveguide with the guide
channel diameter $d$, the solution of the wave equation (\ref{general wave}%
)\ gives for $\vartheta<\vartheta_{c}$ ($\ll1$) \cite{zwanenburg...prl1999}%
\begin{equation}
u_{m}(x)\propto\left\{
\begin{array}
[c]{rcl}%
\sin(k\vartheta_{m}x) & , & \left\vert x\right\vert \leqslant\frac{d}{2}\\
0 & , & \left\vert x\right\vert >\frac{d}{2}%
\end{array}
\right.  \ , \label{1d wave}%
\end{equation}
\noindent where $k_{\parallel m}=k\cos\vartheta_{m}\simeq k\left(
1-\frac{\vartheta_{m}^{2}}{2}\right)  $ with$\ \vartheta_{m}=\frac{\pi
(m+1)}{kd}$. Hence we can define also a maximum number of the waveguide modes
for the fixed wave number $k$ in respect with geometrical parameters of the
waveguide as $m_{\max}=\frac{2d}{\lambda}\vartheta_{c}-1$. From the latter the
condition for a single mode propagation can be reduced as $\lambda_{\perp
}\equiv d$. Hence, as seen from the solution of Helmholtz equation for a
planar guide, we have obtained the same criterion for observing wave features
of radiation propagation as obtained by the phenomenology above presented
(relation (\ref{wave criterium}), where $L\vartheta_{c}\simeq d$, and
$\lambda_{\perp}\simeq\lambda/\vartheta_{c}$).

Situation with a bent planar guide can be considered as a limit case of the
circular guide, analysis of which is given below via example of a capillary system.

\subsection{Circular guide}

\noindent Since in the case of capillary systems, the principal waveguide is a
hollow cylindrical tube (a circular guide), the interaction potential, in
which the waves are propagating, is determined by Eq.(\ref{e-pot-curv})
\noindent with the radiation polarizability parameter $\delta_{0}%
\simeq\vartheta_{c}^{2}/2$ (for simplicity the absorption is considered to be
negligible $\beta_{0}\ll1$). Solving the wave equation for the capillaries
with $\mu$-size holes we are mainly interested in the surface propagation,
which, in fact, defines a wave guiding character inside the channel
($r_{\perp}\simeq r_{1},\ \rho\ll r_{1}$) \cite{alda-nim98,dabagov-xrs2003}%
\[
E_{n}\left(  r\right)  \simeq\sum_{m}C_{m}u_{m}(\rho)\ e^{i\left(
k_{\parallel}z+n\varphi\right)  }\ ,
\]%
\begin{equation}
u_{m}(\rho)\propto\left\{
\begin{array}
[c]{rcl}%
Ai_{m}(\rho) & , & \rho>0\text{ - guiding channel}\\
\alpha{Ai^{\prime}}_{m}(0)\,e^{\alpha\rho} & , & \rho<0\,\,\,\,\,\,(\alpha
>0)\text{ - cladding}%
\end{array}
\right.  \ ,\label{a}%
\end{equation}
\noindent where $Ai_{m}(x)$ is the Airy function, and $\alpha$ is the
arbitrary unit characterizing the capillary substance. Evidently, these
expressions are valid only for the lower-order modes and in the vicinity of a
channel surface. \noindent The expression (\ref{a}) characterizes the waves
that propagate close to the waveguide wall, or in other words, the equation
describes the grazing modal structure of the electromagnetic field inside a
capillary (surface X-ray channeling states). The solution shows also that the
wave functions are damped both inside the channel wall and moving from the
wall towards the center. It should be underlined here that the bound modal
propagation takes place without the wave front distortion that is important
for explaining the interference behind capillary systems (in multiple
reflection optics, \cite{da-ufn2003} and Refs. in).

The analysis of these expressions allows us also to conclude that almost all
radiation power is concentrated in the hollow region and, as a consequence, a
small attenuation along the waveguide walls is observed. However, evaluating
the solution of the wave equation inside the cladding, we can estimate the
tunneling effect (penetration of radiation deep into the cladding), which
becomes significant for the guide acceptance and propagation in both thin wall
and long legth guides .

\noindent\noindent As for the supported modes of the electromagnetic field,
estimating a characteristic radial size of the main grazing mode ($m=0$)
results in
\begin{equation}
2\pi^{2}\overline{u}_{0}^{3}\ \simeq\lambda^{2}r_{1}\quad,
\end{equation}
\noindent and we can conclude that the typical radial size $\overline{u}_{0}$
may overcome the wavelength $\lambda$, whereas the curvature radius $r_{1}$ in
the trajectory plane exceeds the inner channel radius, $r_{0}$: $\overline
{u}_{0}\gg\lambda$ (for example, $\overline{u}_{0}\gtrsim0.1\ \mu$m for a
capillary channel with the radius $r_{0}=10\ \mu$m).

\section{Bulk channeling: $n$-channels}

Above we have considered the transmission of X-ray beams by the guides of
$\mu$-size channels. As shown, in that case we deal with the surface
channeling of radiation due to the fact that the channel sizes are\ much
larger than the radiation wavelength and even than the transverse wavelength.
However, the situation sharply changes in the case when the sizes of channels
become comparable with the radiation transverse wavelength. In practice it
means, that the angle of diffraction for the given wave, determined as
$\vartheta_{d}=\lambda/d$ ( $d$ is the guide transverse size), becomes
comparable with a critical angle of total external reflection; and the
transverse wavelength of a photon approaches the guiding gap: $\lambda_{\perp
}/d\sim1$. In this case, at the specific conditions, channeling of photons
(note, not surface channeling!) in channels of capillary systems may take
place, i.e. actually, in other words, we deal with X-ray waveguiding similar
to light waveguiding in fiber optics. Situation is similar to channeling of
charged particles in crystals, that's why it may be considered as \textit{bulk
channeling} (the quantum well is formed by the bulk properties of a guiding
channel) \cite{ch2004-2006}.

Of course, in this limit the radiation penetration into the cladding becomes
to be important (significant) for resolving propagation problem that speaks on
the necessity of taking into account solution of the wave equation inside the cladding.

\subsection{Planar guide}

\noindent Solving Eq.(\ref{general wave}) for a planar 1D waveguide, we obtain
the amplitudes of channeling states inside the guiding core $\left\vert
x\right\vert \leqslant d$ as
\begin{equation}
E_{m}(x)\propto\left\{
\begin{array}
[c]{rcl}%
\cos(k\vartheta_{m}x) & , & \text{even mode}\\
\sin(k\vartheta_{m}x) & , & \text{odd mode}%
\end{array}
\right.  \ ,\label{1d-Em-core}%
\end{equation}
\noindent and inside the cladding $\left\vert x\right\vert >d$ -
\begin{equation}
\left\vert E_{m}(x)\right\vert \propto e^{-k\sqrt{\vartheta_{c}^{2}%
-\vartheta_{m}^{2}}\left\vert x\right\vert }\ ,\label{1d-Em-cladding}%
\end{equation}
and the dispersion equations for the channeling quantum states -
\begin{equation}
\tan(k\vartheta_{m}d)=\left\{
\begin{array}
[c]{rcl}%
\left(  \frac{\vartheta_{c}^{2}}{\vartheta_{m}^{2}}-1\right)  ^{1/2} & , &
\text{even mode}\\
-\left(  \frac{\vartheta_{c}^{2}}{\vartheta_{m}^{2}}-1\right)  ^{-1/2} & , &
\text{odd mode}%
\end{array}
\right.  \text{ \ }\label{disp_p-guide}%
\end{equation}
The dispersion equations for a planar guide are defining the modal structure
of radiation propagation, i.e. the quantum states of channeling. For instance,
the phase $\Delta\varphi\equiv k\vartheta_{c}d$ determines character of
radiation transmission: if the number of quantum states of channeling is
large, $d/\lambda_{\perp}\gg1$, then we obtain $\Delta\varphi\gg2\pi$ that
speaks on possibility of\ using the ray optics approach for describing the
radiation propagation.

Evidently, there is a special interest to consider a waveguide, which is
formed by the wall of periodic $n$-channels (multilayer planar guides or
narrow multislit system), with a central guiding air gap $d_{0}$ and the
distance $d$ between the layers composing a waveguide wall \cite{buda-nc02}.
For the sake of simplicity, interaction potential in such a waveguide system
may be presented as follows:%
\begin{equation}
V\left(  x\right)  =\sum_{n}V_{n}\left(  x\right)  =k_{\perp}^{2}\left[
1+\Delta\sum_{n}\delta\left(  \left\vert x\right\vert -\frac{d_{0}}%
{2}-nd\right)  \right]  \ ,\label{v-period}%
\end{equation}
\noindent where $\Delta\equiv\overline{\delta}_{0}d$ is the spatially averaged
polarizability of the wall cladding.

\noindent Taking into account the boundary conditions and because of the
potential symmetry, one may conclude that for the central channel $\left\vert
x\right\vert \leq d_{0}/2$, solution of the propagation equation in the
transverse plane will be defined as
\begin{equation}
E_{0}(x)\propto\left\{
\begin{array}
[c]{rcl}%
\cos(k_{\perp}x) & , & \text{even\quad mode}\\
\sin(k_{\perp}x) & , & \text{odd\quad mode}%
\end{array}
\right.  \label{e-central}%
\end{equation}
\noindent Solution for the 1st layer $d_{0}/2\leq\left\vert x\right\vert \leq
d_{0}/2+d$ \ is superposition of the opposite-directed waves
$E(r)=b\ e^{ik_{\perp}x}+c\ e^{-ik_{\perp}x}$. Then, we impose on the
solutions the requirements that the wave function $E$ and its transverse
derivative $E_{x}^{^{\prime}}$ be continuous at the wall-channel boundary
\noindent taking into account the Bloch theorem $E\left(  x+d\right)
=e^{i\kappa d}E(x)$ for the periodical potential function $V\left(  x\right)
=V\left(  x+d\right)  $. From these expressions we obtain the dispersion
relations for even and odd states
\begin{equation}
\left\{
\begin{array}
[c]{c}%
\tan\frac{k_{\perp}d_{0}}{2}\\
\cot\frac{k_{\perp}d_{0}}{2}%
\end{array}
\right\}  =\left\{
\begin{array}
[c]{c}%
-\frac{k^{2}\Delta}{k_{\perp}}+\frac{\cos\left(  k_{\perp}d\right)
-e^{i\kappa d}}{\sin\left(  k_{\perp}d\right)  }\\
\frac{k^{2}\Delta}{k_{\perp}}-\frac{\cos\left(  k_{\perp}d\right)  -e^{i\kappa
d}}{\sin\left(  k_{\perp}d\right)  }%
\end{array}
\right\}  \label{dispers}%
\end{equation}
\noindent that allow the eigenvalue problem to be solved. \noindent Finally
the wave functions of the supported modes for the narrow channel $\left\{
k_{\perp}d_{0},k_{\perp}d\right\}  \ll1$ can be presented by the following
\noindent

$E_{n}(x)\simeq$%
\begin{equation}
\left\{
\begin{array}
[c]{rcl}%
\cos(k_{\perp}x)\ e^{ikz} & , & \left\vert x\right\vert \leq\frac{d_{0}}{2}\\
\cos\frac{k_{\perp}d_{0}}{2}\ \frac{e^{i\kappa x}\sin\left(  k_{\perp
}\left\vert \widetilde{x}\right\vert \right)  -\sin\left[  k_{\perp}\left(
\left\vert \widetilde{x}\right\vert -d\right)  \right]  }{\sin\left(
k_{\perp}d\right)  }\ e^{i\left(  \kappa nd+kz\right)  } & , & \text{n-th
layer}%
\end{array}
\right.  ,
\end{equation}
\noindent where $\left\vert \widetilde{x}\right\vert \equiv\left\vert
x\right\vert -d_{0}/2-nd$, and we see that the Eqs.(\ref{dispers}) may be
solved only for the even modes. However, it is more important to underline
that the even mode exists for any ratio between the channel size and the layer
distance. The spatial distribution of the mode has a maximum at the channel
center, and due to the leak through the potential barrier of wall layers (the
tunneling) we observe the propagation of radiation in the cladding. The
radiation intensity for the successive layer decreases following an
exponential law and is characterized by a local maximum far from the layer
wall. Owing strong tunneling of radiation the acceptance of such a waveguide
system can much exceed the critical angle $\vartheta_{c}$
\cite{beloshitsky...ch2006report}.

\subsection{Circular guide}

Well known solution of the wave equation for a circular guide with the channel
diameter $d$ \ (that corresponds to X radiation propagation in a
$n$--capillary) can be reduced if to write the wave function of
electromagnetic field as$\ E(\mathbf{r}_{\perp})=u(\rho)e^{im\varphi}%
$,\ \ where $\rho$ is the radial coordinate and $\varphi$ is the azimuthal
angle. In this case the wave equation for the radial part of the field
amplitude becomes%
\begin{equation}
\rho^{2}\frac{d^{2}u}{d\rho^{2}}+\rho\frac{du}{d\rho}+\left(  k^{2}%
(\vartheta^{2}-\vartheta_{c}^{2})\rho^{2}-m^{2}\right)
u=0\ ,\label{wave-eq_cyl}%
\end{equation}
\noindent from which the solutions of (\ref{wave-eq_cyl}) in both a hollow
part of the guide and the guide wall cladding -%
\begin{equation}
u(\rho)\propto\left\{
\begin{array}
[c]{rcl}%
J_{m}(k\vartheta_{m}\rho) & , & \rho\leqslant d\\
K_{m}\left(  k\rho\sqrt{\vartheta_{c}^{2}-\vartheta_{m}^{2}}\right)   & , &
\rho>d
\end{array}
\right.  \ ,\label{u_radial}%
\end{equation}
\noindent and the dispersion equations -
\begin{equation}
\frac{\left.  \left(  \partial_{\rho}J_{m}/J_{m}\right)  \right\vert _{\rho
=d}}{\left.  \left(  \partial_{\rho}K_{m}/K_{m}\right)  \right\vert _{\rho=d}%
}=\left(  \frac{\vartheta_{c}^{2}}{\vartheta_{m}^{2}}-1\right)  ^{1/2}%
\label{disp_c-guide}%
\end{equation}
can be obtained.

Here $J_{m}(y)$ and $K_{m}(y)$ are the 1st kind and the modified 2nd kind
Bessel functions, respectively, giving us the bound states for X-ray
channeling in a hollow core of the waveguide. The constants of the
proportionality can be obtained using the matching and normalization
conditions. It is more important to note the asymptotic behaviours of the
Bessel functions specified. Namely, in the guide core we can use $J_{m}%
(\rho)\rightarrow\rho^{-1/2}\cos\alpha\rho$ that proves the modal character of
propagation (different bound channeling states) with the mode amplitudes
decrease from the maximum at the guide center $\rho=0$ to the minimum at the
guide wall $\rho=d$. In the same time we can estimate suppression of the
radiation penetration in the guide cladding for $\rho>d$ as $K_{m}%
(\rho)\rightarrow\rho^{-1/2}e^{-\rho}$. However, being suppressed the
radiation penetrates rather deep in the cladding resulting finally in the
strong tunneling. This phenomenon, first seems to be useless, provides, in
turn, essential increase in the waveguide acceptance.

Speaking on circular guide systems it would be important also to consider
natural nanotube systems (carbon or carbon based nanotubes, alumina porous
membranes) \cite{iij-nat91,saito...book1998}. Due to the nanotube morphology,
i.e. presence of inner cavity, that speaks on possibility of the efficient
transmittance of X-ray, thermal neutron and charged particle beams, a nanotube
can be considered as a capillary of very small inner diameter and wall
thickness \cite{zhevago,dedkov}. However, there is a strong difference between
typical glass capillary and nanotube from the point of view of radiation
propagation through these structures. First of all, the dielectric function as
a function of the distance from the center of glass microcapillary channel
varies by a step-law from zero for the inner hollow cavity to the constant
value defined by the substance, for the channel wall. On the contrary, in case
of nanotube channel, we have continuous change of the dielectric parameter
value. However, the main factor, which defines the character of radiation
propagation inside nanotubes, remains the same as for $n$-channels; it is the
ratio $\lambda_{\perp}/d$.

\noindent There is another very important behaviour to pay attention too.
Because of the small wall thickness of nanotube channels (less than
$\lambda_{\perp}\lesssim100$ \AA ) we have to note that part of the radiation,
channeling inside a nanotube structure, will undergo \textquotedblright
tunneling\textquotedblright\ through the potential wall barrier. A simple
analysis of the radiation propagation in systems both for the case of
macroscopic channel and for the case of totally isotropic spatial structure,
shows the presence of the main channeling mode (the main bound state) for any
structure, whereas the high modes may be suppressed for specific channel
sizes. Hence, nanotubes present a special interest as waveguides, which allow
the supported modes to be governed \cite{childs...phe2003}. Moreover, there is
a special interest in studying the dispersion of radiation in a nanosystem
with a multilayered wall. As follows from the analysis of the general equation
of radiation propagation considered above, at any correlation between the
channel size and the interlayer distance at least one mode (bound state)
should be formed in such a structure. In that case the diffraction of waves
reflected from various layers of the channel wall should be observed, hence
affecting the radiation distribution at the exit of system.

\noindent Evidently, the efficiency of these structures for applications have
to be analyzed, despite the importance of the nanotube X-ray waveguide
phenomenon from the fundamental point of view. \noindent The problems
associated with X-ray and neutron channeling in capillary nanotubes (single-
and multi-wall systems) present a special interest. The first observation of
X-ray channeling in a forest of multiwall carbon nanotubes and successive
analysis of the results based on X-ray channeling theory
\cite{okotrub...jetpl2005,dabagov-aip-proc2003} has shown that this process is
accompanying by X-ray diffraction on the multilayer wall. The latter takes
place due to the strong tunneling of radiation through the nanotube wall cladding.

\section{Resume}

Solution of Maxwell equations describing propagation of electromagnetic waves
in media, where the index of refraction changes as a step function, results in
forming a discrete set of the modes \cite{metzger-science2002}. Presently, the
waveguides for various types radiation are in wide use. Various kinds of them
enable to shape the beams of radiation in different energy ranges that makes
the waveguides to be extremely attractive for applications. For instance,
nowadays it is difficult to imagine any high tech instruments without various
cavities for $\mu$- and radio waves, optical fibers, etc. Among them X-ray
waveguides are mainly in the research stage of development.

Analysis of radiation propagation through the guides of various shapes, above
presented, has shown that all the observed features can be described within an
unified theory of X-ray channeling: \textit{surface channeling in }$\mu
$\textit{-size guides and bulk channeling in }$n$\textit{-size guides}. The
main criterion defining character of radition propagation is the ratio between
the transverse wavelength of radiation and the effective size of a guide, i.e.
$\lambda_{\perp}/d\equiv\vartheta_{d}/\vartheta_{c}$, in other words, the
ratio between the diffraction and Fresnel angles. When this ratio is rather
small, i.e. when the number of bound states is large, the ray optics
approximation is valid. In turn, when $\lambda_{\perp}\simeq d$, a few modes
will be formed in a quantum well; and just a single mode - for $\lambda
_{\perp}\gg d$. Obviously, the latter requires solution of the wave equation
for describing all the features of radiation propagation in such guides. 

Recently, it was shown that at the center of a guide the flux peaking of X
radiation, i.e. the increase of the channeling state intensity at the center
of a guide, should take place \cite{beloshitsky...ch2006report}. This feature
is a proper channeling effect that can be explained only by the modal regime
of radiation propagation, and may find an interesting application for the
purposes of extreme focusing.

It is also important to note that all the considerations taken for X-rays
should be valid for thermal neutrons.

\section*{Acknowledgements}

I am grateful to all my colleagues at LNF\ INFN (Frascati), Laboratory for
High Energy Electrons LPI (Moscow), and Institute for R\"{o}ntgen Optics
(Moscow) actively contributed to the obtained results and supported my work.
The fruitful discussions with M. Kumakhov, V. Beloshitsky, D. Gruev, P.
Childs, S. Kukhlevsky and Yu. Dudchik are specially acknowledged.


\begin{thebibliography}{99}                                                                                               %


\bibitem {spiseg-apl1974}E. Spiller, and A. Segmuller, Appl. Phys. Lett.
\textbf{24 }(1974) 60.

\bibitem {pogossian-opcom1995}S.P.Pogossian, Opt. Comm. \textbf{114} (1995) 235.

\bibitem {da-ufn2003}S.B.~Dabagov, Phys. Uspekhi \textbf{46} (2003) 1053.

\bibitem {kuko-phrep1990}M.A.~Kumakhov, and F.F.~Komarov, Phys. Rep.
\textbf{191} (1990) 289.

\bibitem {Engstrom}P. Engstr\"{o}m, S. Larsson, and A. Rindby, Nucl. Instr.
Meth. \textbf{A302} (1991) 547.

\bibitem {Stern}D.J. Thiel, D.H. Bilderback, A. Lewis, and E.A. Stern, Nucl.
Instr. Meth. \textbf{A317} (1992) 597.

\bibitem {zhevago}N.K. Zhevago, and V.I. Glebov, Phys. Lett. \textbf{A250}
(1998) 360.

\bibitem {dedkov}G.V. Dedkov, Nucl. Instr. Meth. \textbf{B143} (1998) 584.

\bibitem {childs...phe2003}P.A.~Childs, and A.G.~O'Neill, Physica \textbf{E19}
(2003) 153.

\bibitem {zwanenburg...prl1999}M.J.~Zwanenburg, J.F.~Peters, J.H.H.~Bongaerts,
et al., Phys. Rev. Lett. \textbf{82} (1999) 1696.

\bibitem {okotrub...jetpl2005}A.V.~Okotrub, S.B.~Dabagov, A.G.~Kudashov, et
al., JETP Lett. \textbf{81} (2005) 34.

\bibitem {vino1}A. Vinogradov, V. Kovalev, I. Kozhevnikov, and V. Pustovalov,
Sov. Phys. - Tech. Phys. \textbf{30} (1985) 335.

\bibitem {da-rep92}S.B. Dabagov, "Redistribution of X-Rays Traped in Bound
States by Capillary Systems" (Research Report of FIROS: Nalchik-Moscow, 1992).

\bibitem {da...jsr95}S.B. Dabagov, M.A. Kumakhov, S.V. Nikitina, et al., J.
Synchrotron Rad. \textbf{2} (1995) 132.

\bibitem {da...pla1995}S.B. Dabagov, S.V. Nikitina, and M.A. Kumakhov, Phys.
Lett. \textbf{A203} (1995) 279.

\bibitem {daku-spie95}S.B. Dabagov, and M.A. Kumakhov, Proc. SPIE
\textbf{2515} (1995) 124.

\bibitem {alda-nim98}Yu.M. Alexandrov, S.B. Dabagov, M.A. Kumakhov, et al.,
Nucl. Instr. Meth. \textbf{B134} (1998) 174.

\bibitem {arar-phscr98}N. Artemiev, A. Artemiev, V. Kohn, and N. Smolyakov,
Phys. Scripta \textbf{57} (1998) 228.

\bibitem {coh-incoh}S.B. Dabagov, A. Marcelli, V.A. Murashova, et al., Appl.
Opt. \textbf{39} (2000) 3338; S.B. Dabagov, V.A. Murashova, N.L.
Svyatoslavsky, et al., Proc. SPIE. \textbf{3444} (1998) 486.

\bibitem {liu...prl97}Chien Liu, and J.A. Golovchenko, Phys. Rev. Lett.,
\textbf{79} (1997) 788.

\bibitem {kukh.etal-nimb00}S.V. Kukhlevsky, F. Flora, A. Marinai, et al.,
Nucl. Instr. Meth. \textbf{B168} (2000) 276.

\bibitem {ca-da...apl01}G. Cappuccio, S.B. Dabagov, C. Gramaccioni, and A.
Pifferi, Appl. Phys. Lett. \textbf{78} (2001) 2822.

\bibitem {dabagov-xrs2003}S.B. Dabagov, X-Ray Spectrom. \textbf{32} (2003) 179.

\bibitem {zwanenburg...prl2000}M.J.~Zwanenburg, J.H.H.~Bongaerts, J.F.~Peters,
et al., Phys. Rev. Lett. \textbf{85} (2000) 5154.

\bibitem {pfeiffer...science2002}F. Pfeiffer, C. David, M. Burghammer, et al.,
Science \textbf{297} (2002) 230.

\bibitem {bongaerts...jsr2002}J.H.H. Bongaerts, C. David, M. Drakopoulos, et
al., J. Synchrotron Rad. \textbf{9} (2002) 383.

\bibitem {bergemann...prl2003}C. Bergemann, H. Keymeulen, and J.F. van der
Veen, Phys. Rev. Lett. \textbf{91} (2003) 204801.

\bibitem {jarre...prl2005}A. Jarre, C. Fuhse, C. Ollinger, et al., Phys. Rev.
Lett. \textbf{94} (2005) 074801.

\bibitem {fuhse.salditt-phis.b2005}C. Fuhse, and T. Salditt, Physica
\textbf{B357} (2005) 57.

\bibitem {fuhse.salditt-optcomm2006}C. Fuhse, and T. Salditt, Opt. Comm.
\textbf{265} (2006) 140.

\bibitem {bukreeva...prl2006}I. Bukreeva, A. Popov, D. Pelliccia, et al.,
Phys. Rev. Lett. \textbf{97} (2006) 184801.

\bibitem {bilderback-xrs2003}D. Bilderback, X-Ray Spectrom. \textbf{32} (2003) 195.

\bibitem {kukhlevsky-xrs2003}S. Kukhlevsky, X-Ray Spectrom. \textbf{32} (2003) 199.

\bibitem {mono}S.B. Dabagov, and A. Marcelli, Appl. Opt. \textbf{38} (1999) 7494.

\bibitem {Lindhard}J. Lindhard, Kgl. Dan. Vid. Selsk. Mat.-Fys. Medd.
\textbf{34}(14) (1965) 1.

\bibitem {kukhlevsky-ch2006report}S. Kukhlevsky, "Spatiotemporal localization
of coherent and incoherent x-rays by mono and polycapillary optics: from $\mu
$-meter/second to $n$-meter/attosecond", Report to the\ "Channeling 2006"
conference, 3-7 July 2006 (see link \cite{ch2004-2006}).

\bibitem {ch2004-2006}"Channeling 2004" - International Conference on Charged
and Neutral Particles Channeling Phenomena, ed. S. Dabagov, Proc. SPIE Vol.
\textbf{5974 (}http://www.lnf.infn.it/conference/channeling2004); \ 

"Channeling 2006" - http://www.lnf.infn.it/conference/channeling2006.

\bibitem {buda-nc02}E. Burattini, S.B. Dabagov, and F. Monti, Nuovo Cimento
\textbf{117B} (2002) 769.

\bibitem {beloshitsky...ch2006report}V.V. Beloshitsky, D.I. Gruev, and M.A.
Kumakhov, "Theory of penetration of x-rays through glass nano waveguides",
Report to the "Channeling 2006" conference, 3-7 July 2006 (see link
\cite{ch2004-2006}).

\bibitem {iij-nat91}S.Iijima, Nature \textbf{354} (1991) 56.

\bibitem {saito...book1998}R. Saito, G. Dresselhaus, and M. S. Dresselhaus,
\textquotedblright Physical Properties of Carbon Nanotubes\textquotedblright%
\ (Imperial College Press: London, 1998).

\bibitem {dabagov-aip-proc2003}S.B. Dabagov, \textquotedblleft From Surface
down to Bulk X-Ray Channeling\textquotedblright, in book X-Ray and Inner-Shell
Processes, Eds. A. Bianconi, A. Marcelli, and N.L. Saini, AIP Conference Proc.
\textbf{652} (NY, 2003) 89.

\bibitem {metzger-science2002}T.H. Metzger, Science \textbf{297} (2002) 205.
\end{thebibliography}
\end{document}